\documentclass[
amsmath,amssymb,
aps,
prc,
]{revtex4-2}
 
\usepackage{graphicx}% Include figure files
\usepackage{dcolumn}% Align table columns on decimal point
\usepackage{bm}% bold math
\usepackage{CJK}
\usepackage[colorlinks,linkcolor=blue,urlcolor=blue,citecolor=blue]{hyperref}

\usepackage{tikz,xcolor,hyperref}

\definecolor{lime}{HTML}{A6CE39}
\DeclareRobustCommand{\orcidicon}{
	\begin{tikzpicture}
	\draw[lime, fill=lime] (0,0) 
	circle [radius=0.16] 
	node[white] {{\fontfamily{qag}\selectfont \tiny ID}};
	\draw[white, fill=white] (-0.0625,0.095) 
	circle [radius=0.007];
	\end{tikzpicture}
	\hspace{-2mm}
}
\foreach \x in {A, ..., Z}{%
	\expandafter\xdef\csname orcid\x\endcsname{\noexpand\href{https://orcid.org/\csname orcidauthor\x\endcsname}{\noexpand\orcidicon}}
}

\begin{document}
\begin{CJK}{UTF8}{gbsn}

\preprint{APS/123-QED}

\title{$\alpha+\alpha$+${}^{3}$He cluster structure in ${}^{11}$C }% Force line breaks with \\

\author{Ying-Yu Cao (曹颖逾)}
\affiliation{Key Laboratory of Nuclear Physics and Ion-beam Application (MOE), Institute of Modern Physics, Fudan University, Shanghai 200433, China}

\author{De-Ye Tao (陶德晔)}
\affiliation{Key Laboratory of Nuclear Physics and Ion-beam Application (MOE), Institute of Modern Physics, Fudan University, Shanghai 200433, China}

\author{Bo Zhou (周波)\orcidC{}}
\email{zhou\_bo@fudan.edu.cn}
\affiliation{Key Laboratory of Nuclear Physics and Ion-beam Application (MOE), Institute of Modern Physics, Fudan University, Shanghai 200433, China}
\affiliation{Shanghai Research Center for Theoretical Nuclear Physics, NSFC and Fudan University, Shanghai 200438, China}

\author{Yu-Gang Ma (马余刚)\orcidD{}}
\email{mayugang@fudan.edu.cn}
\affiliation{Key Laboratory of Nuclear Physics and Ion-beam Application (MOE), Institute of Modern Physics, Fudan University, Shanghai 200433, China}
\affiliation{Shanghai Research Center for Theoretical Nuclear Physics, NSFC and Fudan University, Shanghai 200438, China}

\date{\today}

\begin{abstract}

We study the $\alpha + \alpha$ + ${}^{3}$He cluster structure of ${}^{11}$C within the microscopic cluster model. The calculations essentially reproduce the energy spectra for both negative and positive parity states, particularly the $3/2_3^-$ state near the $\alpha+\alpha$+${}^{3}$He threshold. We also calculate the isoscalar monopole, electric quadrupole transition strengths, and root-mean-square radii for the low-lying states. These results suggest that the $3/2_3^-$, $1/2_2^-$, and $5/2_3^-$ states have a well-developed $\alpha + \alpha$ + ${}^{3}$He cluster structure. The analysis of the generator coordinate method wave functions indicates the dilute gaslike nature for the $3/2_3^-$, $1/2_2^-$, and $5/2_3^-$ states, suggesting that they could be candidates for the Hoyle-analog state. Furthermore, it is found that the $5/2_2^+$ and $5/2_3^+$ states may possess a linear chain structure.

% \begin{description}
% Secondary publications and information retrieval purposes.
% You may use the \texttt{description} environment to structure your abstract;
% use the optional argument of the \verb+\item+ command to give the category of each item. 
% \end{description}
\end{abstract}

%\keywords{Suggested keywords}%Use show keys class option if keyword
\maketitle

%\tableofcontents
\section{Introduction}

The formation of clustering plays an important role in the structural and dynamical properties of light nuclear systems~\cite{Wildermuth1977, Ono2019PPNP,ikeda1981introduction,freer2018microscopic,von2006nuclear} and provides insights into the fundamental properties of atomic nuclei~\cite{kanada2001structure,zhang2017nuclear,Yang2018,Yang,HuangBS,zhou2014}. 
%%%%
Clustering phenomena are of great importance in various research fields~\cite{MaYG,Chen,CaoRX}. In recent years, considering the effects of clustering, a broad spectrum of studies has been advanced, particularly in the field of relativistic heavy-ion collisions~\cite{MaYG2,ZhangYX}, nuclear reactions~\cite{WangSS,CaoYT} and decay~\cite{Xu,Zhu,Zhou}.
In particular, 
the Hoyle state, characterized by a well-developed $3\alpha$ cluster structure of $\rm{}^{12}C ~(0_2^+)$, is important for nucleosynthesis~\cite{hoyle1954nuclear, mitalas1985unconventional, sun2024} and has attracted considerable interest in nuclear physics~\cite{funaki2003analysis,funaki2005resonance, kanada1998variation,shi2021,zhou2019n}. Recent studies show that this dilute $3\alpha$ gaslike nature can be regarded as nuclear Bose-Einstein Condensate (BEC)~\cite{tohsaki2001alpha}.
According to the $3\alpha$ orthogonality condition model (OCM) calculation~\cite{yamada2005single}, the three $\alpha$ clusters in the $0_2^+$ state are condensed into the $0S$ orbit with a high occupancy of about 70 \%.

The search for Hoyle-analog states has been extended to $N\alpha$ clusters~\cite{ yamada2004dilute, he2014giant, funaki2002description}. According to the $4\alpha$ OCM calculation for $\rm {}^{16}O$~\cite{funaki2008alpha}, the $0_6^+$ state,  above the $4\alpha$ threshold, is considered as a candidate for the $4\alpha$ condensate state. Quite recently, in $\rm {}^{20}Ne$, a state located about 3 MeV above the $5\alpha$ threshold, distinguished by a notable amplitude of the $\rm {}^{16}O~(0_6^+) + \alpha$ structure in Tohsaki-Horiuchi-Schuck-R\"opke (THSR) analyses, shows features of the $5\alpha$ condensate~\cite{adachi2021candidates, zhou20235}.

On the other hand, searching for developed three-cluster states, i.e., the Hoyle-analog state, attracts great interest in both experimental and theoretical research. Recent studies have revealed that~\cite{vas1018} excited states of non-self-conjugate nuclei show clustering compositions involving 
$\alpha$, $\rm {}^{3}He$, triton clusters, and valence neutrons. The $3/2_3^-$ ($E_x=8.56$ MeV) state of $\rm {}^{11}B$, located about 100 keV below the $\alpha$-decay threshold~\cite{nishioka1979structure,descouvemont1996application,kawabata20072alpha+}, has been studied using the $\alpha+\alpha+t$ GCM and antisymmetrized molecular dynamics (AMD)~\cite{yamada2010alpha+,kanada20152,zhou20182}. These investigations revealed a significant enhancement of isoscalar monopole (ISM) transition from the $3/2_1^-$ ground state to $3/2_3^-$ state, with $B(E0;\rm{IS})=147$ $\rm fm^4$, a value comparable to the observed $96 \pm 16$ $\rm fm^4$ and similar to that for the Hoyle state in $\rm{}^{12}C$, $B(E0;\rm{IS})=120 \pm 9$ $\rm fm^4$. This implies that the $3/2_3^-$ state, with  its $(0S)^2_{\alpha}(0S)_{t}$ configuration, possesses a well-developed cluster structure. %Moreover, the $1/2_2^+$ state which is located near the $2\alpha+t$ threshold, has a linear-chain-like structure. 
A further example is the $1/2_2^+$ and $1/2_5^+$ states of $\rm{}^{13}C$, characterized by a $s\mbox{-} \rm wave$ valence neutron
in a $(0S)^3_{\alpha}(0S)_{n}$ configuration, are proposed as the candidates for the Hoyle-analog state~\cite{yamada2015hoyle, chiba2020hoyle}.

As a mirror nucleus of $\rm {}^{11}B$, the $\rm{}^{11}C$ has been the focus of experimental and theoretical investigations in the search for Hoyle-analog states~\cite{descouvemont1987microscopic,descouvemont1990microscopic,descouvemont19957be}.
More recently, some resonances in $\rm {}^{11}C$ have been clearly observed~\cite{yamaguchi2013alpha,dell2020experimental,lombardo2022clustering,li2023cluster,charity2023invariant} in the excitation energy range of 8--14 MeV through the $\rm {}^{7}Be+\alpha$ channel. One of resonances observed around 8.10 MeV~\cite{li2023cluster} is assigned as the head of the $K^{\pi}=3/2^-$ rotational band, suggesting it has an $\alpha+\alpha+\rm{}^{3}He$ cluster structure. In Ref.~\cite{dell2020experimental}, a new excited state, the $5/2^-$ state, was discovered with an excitation energy of 9.36 MeV, and it is characterized by clustering nature.

In recent years, theoretical models such as AMD 
have been used to calculate the Gamow-Teller (GT) transition from the ground state of $\rm {}^{11}B$ to the excited states of $\rm {}^{11}C$~\cite{kanada2007negative,kanada2011cluster}. It was found that the $B[\rm GT;\rm {}^{11}B \to \rm {}^{11}C(3/2_3^-)]$ is small, while other GT transitions are large, which is in good agreement with the observed values~\cite{Kawabata2004}. The abnormally small GT transition indicates a significant difference in isospin symmetry between the ground state of $\rm {}^{11}B$ and $3/2_3^-$ state of $\rm {}^{11}C$.
Additionally, the resonating group method (RGM) has been employed to analyze the energy spectra and resonance state widths~\cite{vasilevsky2018hoyle,vasilevsky2018systematic}. These investigations revealed that the narrow resonances states ($3/2^-$, $5/2^-$, and $5/2^+$) in $\rm {}^{11}B$ and $\rm {}^{11}C$,  characterized by large spatial separation, may be classified as the candidate of Hoyle-analog states.

In this work, we adopt the three-cluster GCM  to search for the Hoyle-analog states in $\rm {}^{11}C$. In Sec.~\ref{theor}, we provide a brief overview of the three-cluster GCM and the computational details. Then we present the results of the GCM and discussion in Sec. ~\ref{result}. The conclusions of this study are summarized in Sec.~\ref{summary}.

\section{THEORETICAL FRAMEWORK}
\label{theor}
\subsection{Microscopic cluster wave function and Hamiltonian}
\begin{figure}[htbp]
\centering
\includegraphics[width=7cm]{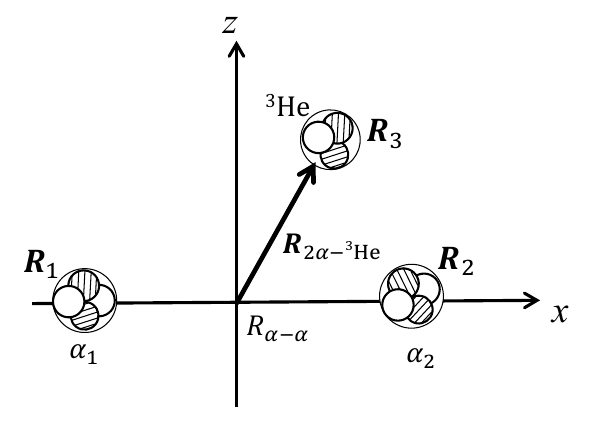}
\caption {Schematic diagram of $\alpha+\alpha+\rm{}^{3}He$ structure of the Brink wave function.}
\label{fig1:11C_configuration}
\end{figure}

The Brink wave function~\cite{brink1965alpha}
 is used as the basis wave function for the GCM calculations of $\rm {}^{11}C$, with the $\rm \alpha+\alpha+{}^{3}He$ cluster configuration shown in Fig.~\ref{fig1:11C_configuration}.

The Brink wave function is defined as follows:
 
\begin{equation}
\begin{aligned}
\label{eq1}
&\Phi(\boldsymbol{R}_1,\boldsymbol{R}_2,\boldsymbol{R}_3)=\mathcal{A}\{\Phi_{\alpha}(\boldsymbol{R}_1)\Phi_{\alpha}(\boldsymbol{R}_2)\Phi_{\rm {}^{3}He}(\boldsymbol{R}_3)\},\\
&\Phi_{\alpha}(\boldsymbol{R})=\mathcal{A}\{\prod_{i=1}^{4}\phi(\boldsymbol{R},\boldsymbol{r}_i)\chi_i \tau_i\},\\
&\Phi_{\rm{}^{3}He}(\boldsymbol{R})=\mathcal{A}\{\prod_{i=1}^{3}\phi(\boldsymbol{R},\boldsymbol{r}_i)\chi_i \tau_i\},\\
&\phi(\boldsymbol{R},\boldsymbol{r}_i)=(\frac{1}{\pi b^2})^{3/4}e^{-\frac{(\boldsymbol{r}_i-\boldsymbol{R})^2}{2b^2}},
\end{aligned}
\end{equation}
where $\Phi_{\alpha}$ and $\Phi_{\rm {}^{3}He}$ denote the wave functions of $\alpha$ and $\rm {}^{3}He$ clusters, respectively. $\boldsymbol{R}_1$, $\boldsymbol{R}_2$, and $\boldsymbol{R}_3$ are the generator coordinates of two $\alpha$ and $\rm {}^{3}He$ clusters, abbreviated as $\{\boldsymbol{R}\}=\{\boldsymbol{R}_1, \boldsymbol{R}_2, \boldsymbol{R}_3 \}$. The condition $(4\boldsymbol{R}_1+4\boldsymbol{R}_2+3\boldsymbol{R}_3)/11=0$ is applied.
$\phi(\boldsymbol{R},\boldsymbol{r}_i)\chi_i \tau_i$ represents the $i$th single-nucleon wave function, where $\phi(\boldsymbol{R},\boldsymbol{r}_i)$ is the spatial wave function. $\chi_i$ and $\tau_i$ denote the spin and isospin of each nucleon. For the $\rm {}^{3}He$ cluster,
the spin of two protons are set to be up and down, respectively, and the spin of the remaining neutron is set to be up. The oscillator width $b=$ 1.46 fm is chosen for all clusters to avoid spurious center-of-mass problems in this work.

The GCM wave functions of $\rm {}^{11}C$ are obtained by superposing over various $\rm \alpha+\alpha+{}^{3}He$ configurations 
\begin{equation}
\begin{aligned}
\label{eq2}
\Psi_M^{J \pi}=\sum_{\{\boldsymbol{R}\}K}f_{\{\boldsymbol{R}\}K} \Phi_{MK}^{J \pi}(\{\boldsymbol{R}\}),
\end{aligned}
\end{equation}
where $\Phi_{MK}^{J \pi}(\{\boldsymbol{R}\})$ is the projected Brink wave functions,
\begin{equation}
\begin{aligned}
\label{eq3}
\Phi_{MK}^{J \pi}(\{\boldsymbol{R}\})=P_{MK}^JP^{\pi}\Phi(\{\boldsymbol{R}\}).
\end{aligned}
\end{equation}
$P_{MK}^J$ and $P_{\pi}$ denote the angular-momentum and parity projector, respectively. $K$ are chosen as: $K=-J,-J+1,\dots,J-1,J$. The coefficients $f_{\{\boldsymbol{R}\} K}$ are determined by solving the Hill-Wheeler equation~\cite{ring2004nuclear}
\begin{equation}
\begin{aligned}
\label{eq4}
\sum_{\{\boldsymbol{R}'\}K'}&f_{\{\boldsymbol{R}'\}K'} \bigg[ \langle \Phi_{MK}^{J \pi}(\{\boldsymbol{R}\})
|\hat{H}|
\Phi_{MK'}^{J \pi}(\{\boldsymbol{R}'\}) \rangle -\\
&E\langle \Phi_{MK}^{J \pi}(\{\boldsymbol{R}\})
|\Phi_{MK'}^{J \pi}(\{\boldsymbol{R}'\}) \rangle
\bigg]=0.
\end{aligned}
\end{equation}

The Hamiltonian employed in this study is as follows,

\begin{equation}
\begin{aligned}
\label{eq5}
\hat{H}=\sum_{i=1}^{11} t_i - t_{\rm{c.m.}}+\sum_{i<j}^{11} V_{ij}^{\rm NN}+\sum_{i<j}^{11} V_{ij}^{C},
\end{aligned}
\end{equation}
where $t_i$  denotes the kinetic energy for the $i$th nucleon, while $t_{\rm c.m.}$ corresponds to the kinetic energy for the center-of-mass. We adopt the Volkov No.2 interaction~\cite{volkov1965equilibrium} combined with the G3RS potential~\cite{yamaguchi1979effective,okabe1979structure} as the nucleon-nucleon interaction, which is given as 
\begin{equation}
\begin{aligned}
\label{eq6}
V_{ij}^{\rm NN}=&\sum_{n=1}^2 V_ne^{-r_{ij}^2/a_n^2}(W+BP_{\sigma}-HP_{\tau}-MP_{\sigma}P_{\tau}) \\
&+\sum_{n=1}^2 w_n e^{-b_nr_{ij}^2} P({}^{3}O)\boldsymbol{L} \cdot \boldsymbol{S}, 
\end{aligned}
\end{equation}
where $P_{\sigma}$ and $P_{\tau}$ stand for the spin and isospin exchange operators, respectively. $V_1 = - 60.65$ MeV, $V_2$ = 61.14 MeV, $a_1$ = 1.80 fm, $a_2$ = 1.01 fm, $W = 1 - M$, $M$ = 0.59, and $B=H=0.125$. For the spin-orbit force, $b_1 = 5.0 \rm \ fm^{-2}$, $b_2 = 2.778 \rm \ fm^{-2}$, and $w_1=-w_2$ = 2200 MeV. $P({}^{3}O)$ represents the projection operator to the triplet-odd states.

\subsection{Overlap between GCM and projected Brink wave functions}
The GCM wave functions are obtained by superposing many different configurations of $\rm \alpha+\alpha+{}^{3}He$, each with distinct generator coordinates. To evaluate the component of some specific $\rm \alpha+\alpha+{}^{3}He$ configurations within the GCM wave functions, we calculate the overlap between the GCM wave functions and projected Brink wave functions for our analysis.

To simplify the calculation of the overlap function, we define the distance between $\alpha_1$ and $\alpha_2$ as $R_{\alpha\text{-}\alpha}$, and the distance between the center of the two $\alpha$ clusters and $\rm {}^{3}He$ as $R_{2\alpha\text{-}{\rm{}^{3}He}}$. Additionally, $R_{(2\alpha\text{-}{}^3\mathrm{He})x}$ and $R_{(2\alpha\text{-}{}^3\mathrm{He})z}$ represent the projections of $R_{2\alpha\text{-}\rm {}^{3}He}$ on the x-axes and z-axes, respectively, as shown in Fig.~\ref{fig1:11C_configuration}. We denote the Brink wave functions for these specific configurations as $\Phi(R_{\alpha\mbox{-}\alpha}, R_{2\alpha\mbox{-}{ }^3 \mathrm{He}})$. Consequently, the overlap between the GCM wave functions and the Brink wave functions of a particular configuration is expressed as follows, 
\begin{equation}
\begin{aligned}
\label{eq7}
O(R_{ \alpha\mbox{-}\alpha}, R_{2\alpha\mbox{-}{ }^3 \mathrm{He}})=&\sum_{KK'}\langle \Psi_M^{J\pi}|P_{MK}^{J\pi}\Phi(R_{\alpha\mbox{-}\alpha}, R_{2\alpha\mbox{-}{ }^3 \mathrm{He}})\rangle 
S_{KK'}^{-1} \\
&\times \langle P_{MK'}^{J\pi}\Phi(R_{\alpha\mbox{-}\alpha}, R_{2\alpha\mbox{-}{ }^3 \mathrm{He}})| \Psi_M^{J\pi} \rangle, 
\end{aligned}
\end{equation}
where $S_{KK'}$ denotes the overlap between the Brink wave functions with projections $K$ and $K'$, 
\begin{equation}
\begin{aligned}
\label{eq8}
S_{KK'}=\langle P_{MK}^{J\pi} \Phi(R_{\alpha\mbox{-}\alpha}, R_{2\alpha\mbox{-}{ }^3 \mathrm{He}})| P_{MK'}^{J\pi} \Phi(R_{\alpha\mbox{-}\alpha}, R_{2\alpha\mbox{-}{ }^3 \mathrm{He}}) \rangle.
\end{aligned}
\end{equation}

\subsection{Isoscalar monopole, dipole, and electric quadrupole transitions}

The ISM transition strength is an important probe of cluster states~\cite{yamada2008monopole,yamada2012isoscalar}. The ISM transition operator is defined as follows:
\begin{equation}
\begin{aligned}
\label{eq9}
\mathcal{M}^{\rm ISM}=\sum_{i=1}^{11}(\boldsymbol{r}_i-\boldsymbol{r}_{\rm {c.m.}})^2, 
\end{aligned}
\end{equation}

The isoscalar dipole (ISD) transition is a good tool for studying asymmetric cluster structures~\cite{chiba2016isoscalar,kanada2016isovector,chiba2017inversion}. The operator and transition strength are defined as follows,  
\begin{equation}
\begin{aligned}
\label{eq10}
\mathcal{M}^{\rm ISD}=\sum_{i=1}^{11}(\boldsymbol{r}_i-\boldsymbol{r}_{\rm{c.m.}})^3Y_{1\mu}(\widehat{\boldsymbol{r}_i-\boldsymbol{r}_{\rm{c.m.}}}), 
\end{aligned}
\end{equation}

\begin{equation}
\begin{aligned}
\label{eq11}
B(\rm {ISD}; J_i \to J_f) = \sum_{M_f, \mu}|\langle J_f M_f|\mathcal{M}^{ISD}_\mu|J_i M_i \rangle|^2, 
\end{aligned}
\end{equation}
where $J_i$ and $J_f$ are the angular momenta of the initial state $|J_i M_i\rangle$ and final state $|J_f M_f\rangle$, respectively. $\boldsymbol{r}_i$ and $\boldsymbol{r}_{\rm{c.m.}}$ denote the coordinate of the $i$th nucleon and the center of mass of the system, respectively. The $Y_{1\mu}$ represents the solid spherical harmonics of degree 1 and order 
$\mu$.

The electric dipole and quadrupole transition strengths $B(E1)$ and $B(E2)$ are important observational quantities, which can be represented as follows,
\begin{equation}
\begin{aligned}
\label{eq12}
\mathcal{M}^{E\lambda}_\mu=e^2 \sum_{i=1}^{11} \frac{1-\tau_{iz}}{2}(\boldsymbol{r}_i-\boldsymbol{r}_{\rm{c.m.}})^2Y_{\lambda\mu}(\widehat{\boldsymbol{r}_i-\boldsymbol{r}_{\rm{c.m.}}}),\ \lambda=1,2 \ , 
\end{aligned}
\end{equation}

\begin{equation}
\begin{aligned}
\label{eq13}
B(E\lambda; J_i \to J_f) = \sum_{M_f, \mu}|\langle J_f M_f|\mathcal{M}^{E\lambda}_\mu|J_i M_i \rangle|^2,\ \lambda=1,2 \ ,
\end{aligned}
\end{equation}
where $\tau_{iz}$ is the isospin projection of the $i$th nucleon.

\section{RESULTS AND DISCUSSIONS}
\label{result}

We perform GCM calculations by superposing 500 $\alpha+\alpha+\rm {}^{3}He$ Brink wave functions. In GCM calculations, the choice of the basis wave functions is crucial to avoid issues arising from the non-orthogonal nature of the basis and potential over-completeness. To construct the basis, we employed two approaches:

First, 284 basis wave functions were generated by sampling the coordinates ${R_{\alpha-\alpha}}$, $R_{(2\alpha-{}^{3}\mathrm{He})_x}$, and $R_{(2\alpha-{}^{3}\mathrm{He})_z}$ from a normal distribution with a mean of $\mu = 0$ and a standard deviation of $\sigma = 2.5$ fm. This approach allows for a broad and randomized exploration of the model space.
Second, an additional set of 216 wave functions was generated using a systematic grid-based method. Specifically, the inter-$\alpha$ particle distance $R_{\alpha-\alpha}$ was varied from 1.5--9 fm in steps of 1.5 fm. The distance between the $^3\mathrm{He}$ particle and the center of the two alpha particles, $R_{(2\alpha-^3\mathrm{He})}$, was varied over the same range and step size. Additionally, the angle between the $^3\mathrm{He}$ particle and the $x$ axis was varied from $\pi/12$ to $\pi/2$ in steps of $\pi/12$. The combination of these parameters resulted in $6 \times 6 \times 6 = 216$ wave functions.
In total, 500 spatial basis wave functions were constructed by combining these two sets. It should be noted that we also superpose different $K$ values ( $-J,-J+1,\dots ,J-1,J$) in our GCM calculations, as shown in Eq.~(\ref{eq4}). In this case, the total number of wave functions for the angular momentum $J$ is, $(2J+1)\times500$. 
To ensure the effectiveness of the basis, we diagonalized the norm kernel and discarded wave functions with smaller eigenvalues (e.g., smaller than 0.001), thereby eliminating poorly constructed or overly redundant basis vectors. These eigenvectors were then used to construct new basis wave functions for the diagonalizations of the Hamiltonian. See details in Ref.~\cite{ring2004nuclear}. This procedure ensured numerical stability and reliability in solving the Hill-Wheeler equation.

\begin{figure*}[htbp]
\centering
\includegraphics[width=18cm]{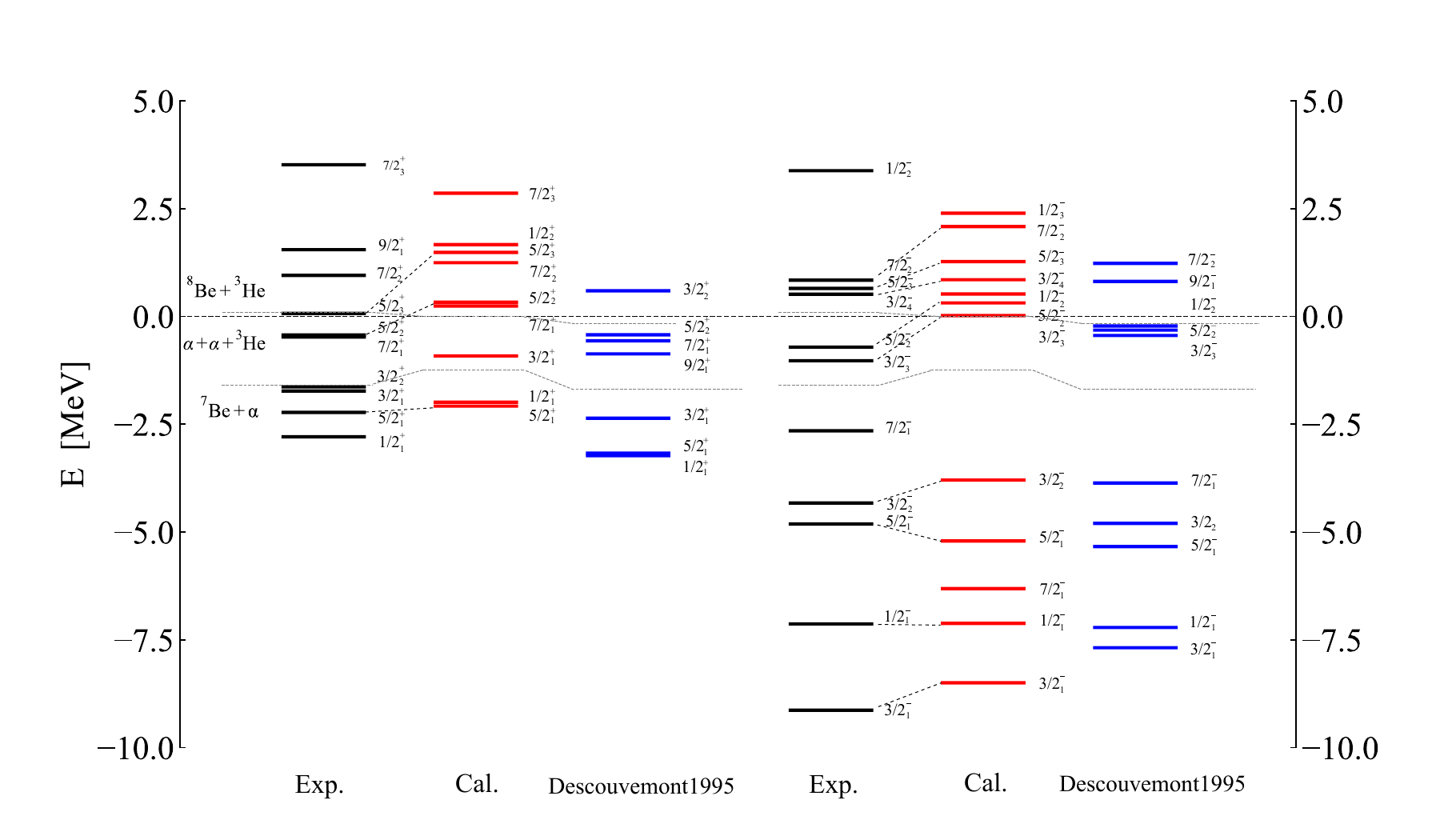}
\caption {Energy levels for $\rm {}^{11}C$ from experimental data~\cite{KELLEY201288}, theoretical calculations from ' Descouvemont1995' (Descouvemont 1995: ~\cite{descouvemont19957be}), and the present calculation are compared. The black dashed line at 0 MeV represents the $\alpha+\alpha+\rm {}^{3}He$ threshold. The two gray dashed lines, from top to bottom, represent the $\rm {}^{8}Be+{}^{3}He$ and $\rm {}^{7}Be+\alpha$ thresholds, respectively.}
\label{fig2:E_level}
\end{figure*}

The energy spectra are shown in Fig.~\ref{fig2:E_level}, compared with the experimental data and the theoretical calculation from Ref.~\cite{descouvemont19957be}. It can be seen that our calculations basically reproduce the
energy spectra for both negative and positive parity states, particularly for $3/2_3^-$ and $5/2_2^+$ states close to the $\alpha+\alpha+\rm {}^{3}He$ threshold. For the negative-parity states, we obtained four $3/2^-$ states ($3/2_1^-,3/2_2^-,3/2_3^- $ and $3/2_4^-$), and the energy gaps are consistent with those observed in experiments. 
The excitation energies of the $3/2_i^- \ (i=1,2,3,4)$ states, obtained from both our calculations and experimental data, are presented in Table~\ref{table_energy}.

\begin{table}[h]
\caption{Excitation energies (units: MeV) of the $3/2_i^- \ (i =1, 2, 3, 4)$ states  from both experimental data~\cite{soic2004alpha, soic2005three, li2023cluster} and theoretical calculations.}
\label{table_energy}
\setlength{\tabcolsep}{30pt} % 
\renewcommand{\arraystretch}{1} % 
\centering
\begin{tabular}{lll} %
\toprule
$J^{\pi}$ & Exp. & Cal. \\ \hline
$3/2_1^-$ & 0.0 & 0.0 \\ 
$3/2_2^-$ & 4.80 & 4.70 \\ 
$3/2_3^-$ & 8.10 & 8.51 \\ 
$3/2_4^-$ & 9.65 & 9.35 \\ 
\toprule
\end{tabular}
\end{table}

Notably, the calculated excitation energy (8.51 MeV) of the $3/2_3^-$ state lies slightly above the $\alpha+\alpha+\rm {}^{3}He$ threshold (8.46 MeV). 
Experimental studies~\cite{soic2004alpha, soic2005three, li2023cluster} have identified excited states in $\rm {}^{11}C$ that decay into $\alpha+\rm {}^{7}Be(3/2^-)$ within the energy range from 8.5--13.5 MeV. It is known that the ground state of the $\rm {}^{7}Be$ have strong $\alpha$ decay involving the $\alpha+\rm {}^{3}He$ channel. In this case, they proposed that the $3/2_3^-$ state may exhibit a three-cluster structure.

On the other hand,  the recent experimental study~\cite{dell2020experimental} found a new excited state ($5/2^-$) at an excitation energy of 9.38 MeV, which shows clustering characteristics. In our calculations, we obtain the $5/2_3^-$ state at $E_x=9.77$ MeV and it might be corresponding to the observed state. The properties and structure of this state will be discussed later. In the positive-parity spectra, we observe a level inversion,  where the calculated $5/2_1^+$ state lies below the $1/2_1^+$ state, contradicting the observed energy level ordering~\cite{yamada2010alpha+}.

Our spectra show some differences compared to the theoretical calculations in Ref.~\cite{descouvemont19957be}, particularly for the $7/2^-$ and $5/2_3^-$ states. The model configuration for $\rm{}^{11}C$ in Ref.~\cite{descouvemont19957be} incorporates two coupling schemes: $\rm {}^{7}Be+\alpha$ and $\rm {}^{8}Be+{}^{3}He$. In their model, the $\rm {}^{8}Be$ wave functions are specified with an $\alpha$-$\alpha$ distance of 3.8 fm, and the $\rm {}^{7}Be$ wave functions an $\alpha$-$\rm {}^{3}He$ distance of 3.7 fm. In contrast, our model space includes these configurations along with random configurations, leading to the observed differences in the calculated spectra.

From the above energy spectra, we obtain several states close to the $\rm \alpha+\alpha+{}^{3}He$ threshold. To clarify their structure, we calculate their root-mean-square (r.m.s.) radii.
For the ground state $3/2_1^-$ of $\rm {}^{11}C$, the nuclear r.m.s. radius is determined to be 2.42 fm. Meanwhile, the calculated nuclear r.m.s.\ radius 
 for the $3/2_3^-$ state is calculated as 3.54 fm, which is comparable to that of the Hoyle state~\cite{PhysRevC.80.064326, PhysRevC.92.021302, PhysRevLett.98.032501}. According to Ref.~\cite{vasilevsky2018hoyle}, the radii for the $3/2_1^-$ and $3/2_3^-$ states of $\rm {}^{11}C$ are obtained as 2.17 fm and 3.23 fm, respectively, which are smaller than our estimates. Moreover, in our results,
 the r.m.s. radii for the $1/2_2^-$, $5/2_2^-$, and $5/2_3^-$ states, which are close to the $\alpha+\alpha+\rm {}^{3}He$ threshold, have been calculated to be 3.58 fm, 2.80 fm and 3.45 fm, respectively.

\begin{figure}[htbp]
\centering
\includegraphics[width=9cm]{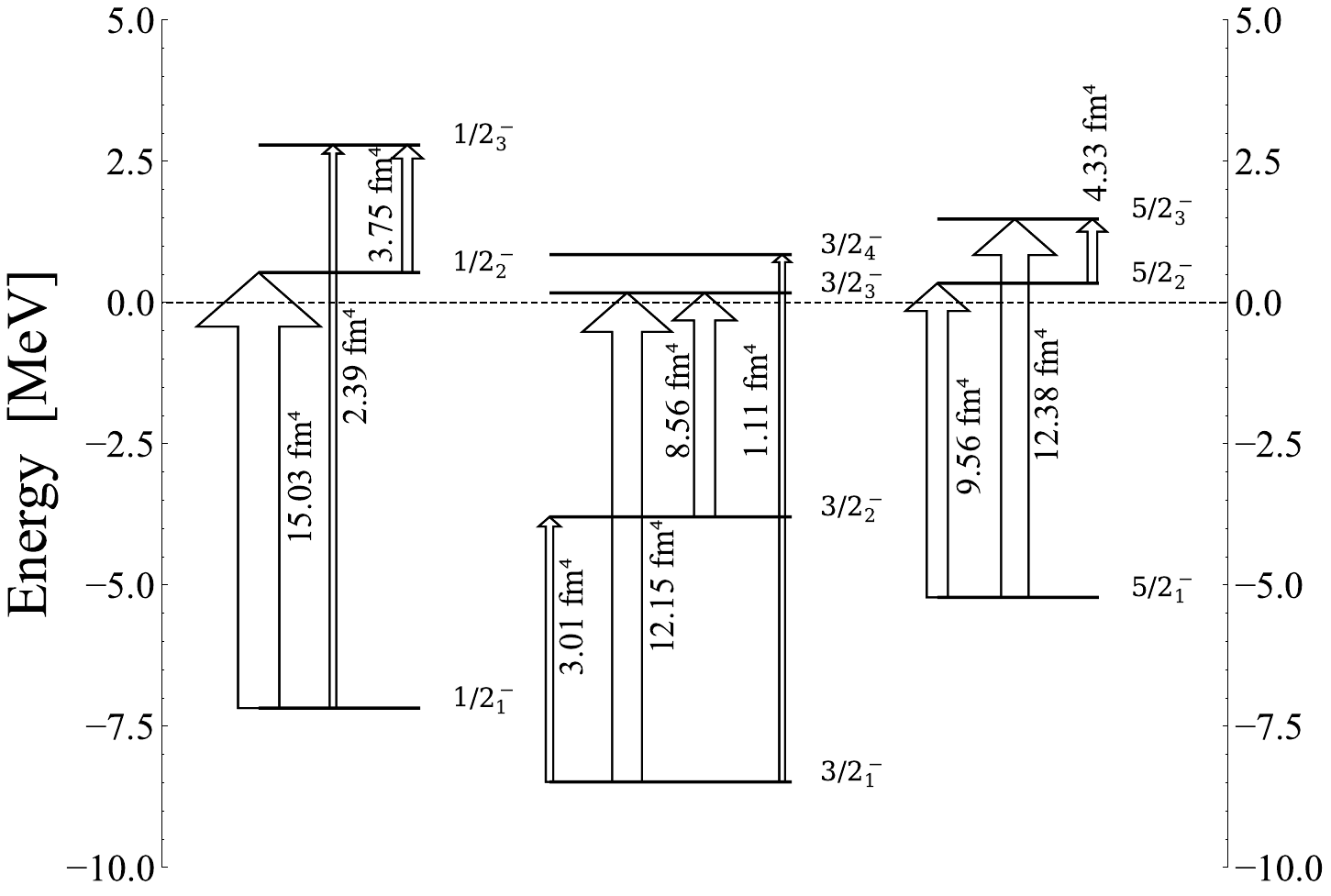}
\caption {The calculated isoscalar monopole transition
strengths for the negative-parity states ($J^\pi=1/2^-$, $3/2^-$, and $5/2^-$ states) in $\rm {}^{11}C$. The dashed line represents the $\alpha+\alpha+\rm {}^{3}He$ threshold.}
\label{fig3:ISM_transition}
\end{figure}

The ISM transition strength is considered a good tool for probing cluster states, as transitions from the ground state to the cluster state often exhibit significant enhancement. The ISM transition operator is given in Eq.~(\ref{eq9}). The ISM transition strengths for the $1/2^-$, $3/2^-$ and $5/2^-$ states are shown in Fig.~\ref{fig3:ISM_transition}. Note that the ISM transition strengths for the $3/2_3^-$, $1/2_2^-$ and $5/2_3^-$ states are considerably enhanced, with
$B(\rm{ISM}; 3/2_1^- \to 3/2_3^- ) = 12.15$ $\rm fm^ 4$, $B(\rm{ISM}; 1/2_1^- \to 1/2_2^- ) = 15.03$ $\rm fm^ 4$, and $B(\rm {ISM}; 5/2_1^- \to 5/2_3^- ) = 12.38$ $\rm fm^ 4$.
Therefore, the $1/2_2^-$, $3/2_3^-$, and $5/2_3^-$ states are close to the $\alpha+\alpha+\rm{}^{3}He$ threshold and are characterized by large r.m.s.\ radii and ISM transition strengths, indicating their clustering structure~\cite{kanada2007negative}. 

ISD transition strength serves as another effective tool for probing clustering. According to Ref.~\cite{chiba2016isoscalar}, asymmetric cluster states are strongly excited by ISD transition. For the positive parity states, the $5/2^+$ states show large radius and ISM transition strengths. We expect that the ISD transition strengths for $5/2^+$ states can be enhanced as well as ISM transition strengths. However, the calculated ISD transition strengths, $B(\rm{ISD}; 3/2_1^- \to 5/2_1^+ ) = 2.33$ $\rm fm^ 6$ and $B(\rm{ISD}; 3/2_1^- \to 5/2_2^+ ) = 3.12$ $\rm fm^ 6$, do not show significant enhancement.

The $B(E2)$ transition strength is used to investigate band structure~\cite{yamaguchi2013alpha} and the deformation of the atomic nuclei. 
We present $B(E2)$ transition strengths for some low-lying states in Table.~\ref{E2transition}. Notably, states with compact shell-model structure, including $1/2_1^-$, $3/2_1^-$ and $5/2_1^-$, show enhancement in their $B(E2)$ transitions strengths.
 The calculated $B(E2, 5/2_1^- \to 3/2_1^-)$ is 9.1 $ e^2 \rm fm^4$ , in close agreement with the 6.8 $ e^2 \rm fm^4$ obtained using the AMD model ~\cite{kanada1997opposite}.

For the low-lying states, the calculated $B(E2)$ transition strengths are $B(E2, 1/2_1^- \to 3/2_1^-) = 13.9$ $e^2 \rm fm^4$ and $B(E2, 3/2_1^- \to 5/2_1^-) = 13.6$ $e^2 \rm fm^4$, which are quite similar. We speculate that the $1/2_1^-$, $3/2_1^-$, and $5/2_1^-$ states may belong to the same rotational band. On the other hand,  
experimental observations have provided information on the negative-parity rotational band ($K^{\pi} = 3/2^-$) in $\rm {}^{11}C$, with the band head corresponding to the $3/2^-_3$ state~\cite{soic2004alpha, soic2005three, yamaguchi2013alpha}. Recent experimental findings further suggest that the third $5/2^-$ state ($E_x=9.38$ MeV) is likely the second member of the $K^{\pi}=3/2^-$ rotational band~\cite{li2023cluster}. Our calculations, which yield radii of 3.54 fm for the $3/2^-_3$ state and 3.45 fm for the $5/2^-_3$ state, along with the closely matching ISM transition strengths, $B(\rm{ISM}; 3/2_1^- \to 3/2^-_3) = 12.15$ $\rm fm^4$ and $B(\rm{ISM}; 5/2_1^- \to 5/2^-_3) = 12.38$ $\rm fm^4$, suggest that these two states share a similar cluster structure and likely belong to the $K^{\pi}=3/2^-$ rotational band. However, due to $K$ mixing, a more detailed discussion of the rotational band is not possible at this time.

\begin{table}[h]
\caption{$E2$ transition strength for the negative-parity states in $\rm {}^{11}C$. (units: $e^2 \rm fm^4$).}
\label{E2transition}
\setlength{\tabcolsep}{30pt} % 
\renewcommand{\arraystretch}{1} % 
\centering
\begin{tabular}{ll} % 
\toprule
$J_i^\pi \to J_f^\pi$       & $B(E2)$   \\  \hline
$1/2_1^- \to 3/2_1^-$       & 13.9 \\
$1/2_1^- \to 3/2_2^-$       & 10.0 \\
$3/2_1^- \to 5/2_1^-$       & 13.6 \\
$3/2_1^- \to 3/2_2^-$       & 2.0  \\
$3/2_1^- \to 3/2_3^-$       & 2.8  \\
$3/2_1^- \to 5/2_2^-$       & 2.9  \\
$5/2_1^- \to 3/2_1^-$       & 9.1  \\
$5/2_1^- \to 5/2_2^-$       & 2.4  \\
\toprule
\end{tabular}
\end{table}

To analyze the $\alpha + \alpha + \rm {}^{3}He$ clustering configuration, we calculate the overlap between the GCM and single projected Brink wave functions, denoted as $O(R_{\alpha\text{-}\alpha}, R_{2\alpha\text{-}{}^3\mathrm{He}})$, as formulated in Eq.~(\ref{eq7}). This overlap reveals the significance of the spatial configuration of the clustering components. We adjust $R_{\alpha\text{-}\alpha}$ to its optimal value and generate contour plots to visualize the overlap $O(R_{\alpha\text{-}\alpha}, R_{2\alpha\text{-}{}^3\mathrm{He}})$ as a function of $R_{2\alpha\text{-}{}^3\mathrm{He}}$. The parameters $R_{(2\alpha\text{-}{}^3\mathrm{He})x}$ and $R_{(2\alpha\text{-}{}^3\mathrm{He})z}$ are uniformly distributed over a range from 0--5 fm, with a mesh size of 0.1 fm for each parameter.

\begin{figure*}[htbp]
\centering
\includegraphics[width=21cm]{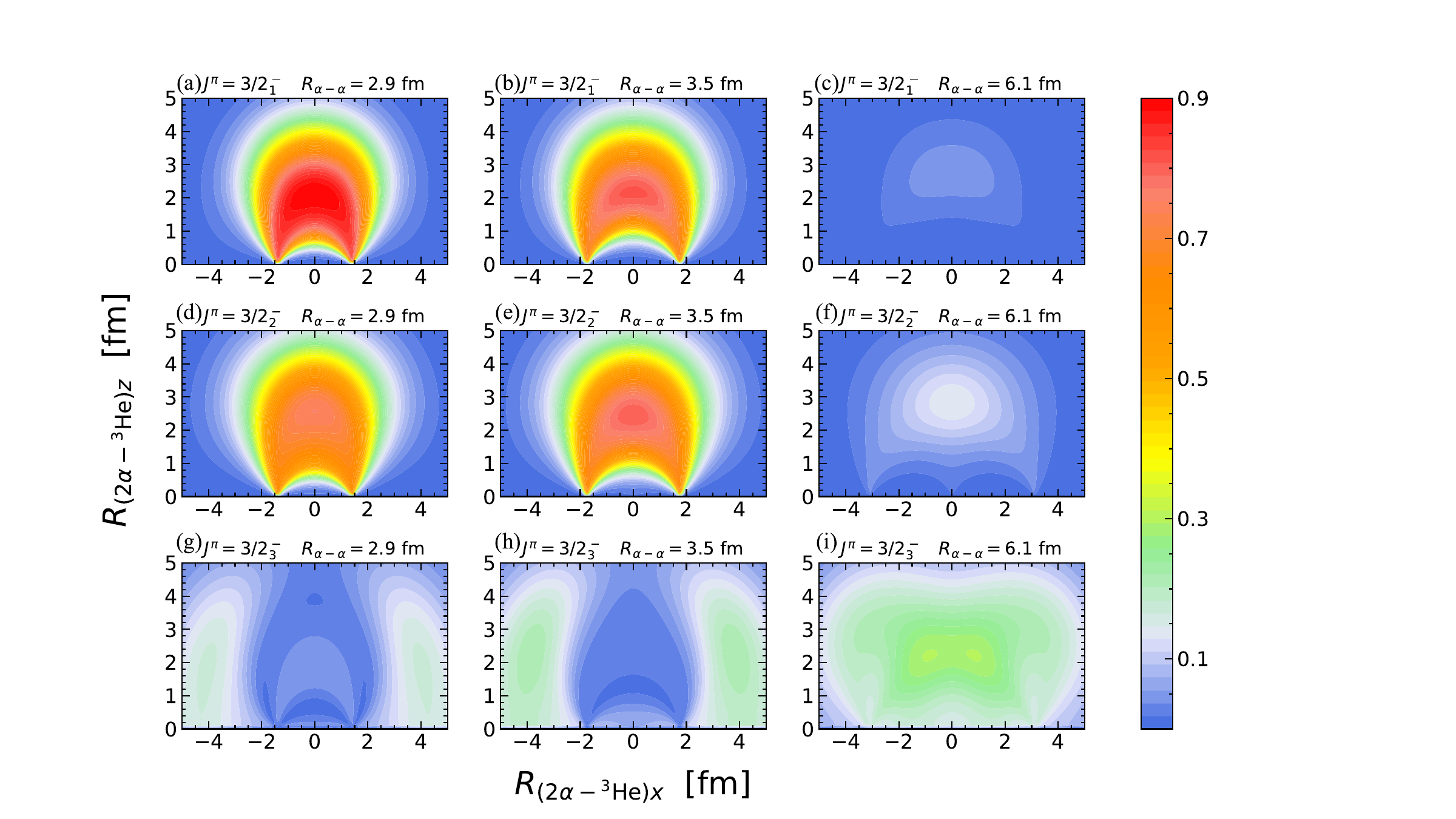}
\caption {Contour plots of the overlap $O(R_{ \alpha\mbox{-}\alpha}, R_{2\alpha\mbox{-}{ }^3 \mathrm{He}})$ for the $J^\pi =3/2^-$ states are presented as a function of the $x$ component and $z$ component of $R_{ {\rm2\alpha\mbox{-}{}^{3}He}}$. (a)--(c) correspond to the overlap for the $3/2_1^-$ state with the distance $R_{ \alpha\mbox{-}\alpha}$  set to $2.9$, $3.5$, and $6.1$ fm, respectively. (d)--(f) for the $3/2_2^-$ state and (g)--(i) for the $3/2_3^-$ state are similarly set to the distances of $2.9$, $3.5$, and $6.1$ fm, respectively.}
\label{fig4:overlap_neg1.5}
\end{figure*}

In Fig.~\ref{fig4:overlap_neg1.5}, we display contour plots of overlaps for the negative-parity states of $\rm {}^{11}C$, specifically the $3/2_1^-$, $3/2_2^-$ and $3/2_3^-$ states. 
Figures 4(a)--4(c) show the change of the overlap with different distances of two $\alpha$ for the $3/2_1^-$ state. From Fig. 4(a), we observe the maximum normalized overlap $\langle \Phi_{3/2^-_1}(R_{\alpha\text{-}\alpha}=2.9~ {\rm fm}, R_{(2\alpha\text{-}{}^3\mathrm{He})z}=2.2~ \rm fm)|\Psi_{\rm GCM} \rangle=0.90$. As the inter-$\alpha$ distance increases to 3.5 fm and 6.1 fm, the values of overlap are $\langle \Phi_{3/2^-_1}(R_{\alpha\text{-}\alpha}=3.5~ {\rm fm}, R_{(2\alpha\text{-}{}^3\mathrm{He})z}=2.6~ \rm fm)|\Psi_{\rm GCM} \rangle=0.81$ and $\langle \Phi_{3/2^-_1}(R_{\alpha\text{-}\alpha}=6.1~ {\rm fm}, R_{(2\alpha\text{-}{}^3\mathrm{He})z}=2.7~ \rm fm)|\Psi_{\rm GCM} \rangle=0.05$, respectively.
Therefore, the obtained maximum overlap with $(R_{\alpha\text{-}\alpha}=2.9~ {\rm fm}, R_{(2\alpha\text{-}{}^3\mathrm{He})z}=2.2~ \rm fm) $ indicates that the $3/2_1^-$ state has a compact spatial distribution and conforms to the shell-model-like structure.

Figures 4(d)--4(f) show the inter-$\alpha$ distance of 2.9, 3.5 and 6.1 fm, with the normalized overlaps being $\langle \Phi_{3/2^-_2}(R_{\alpha\text{-}\alpha}=2.9~ {\rm fm}, R_{(2\alpha\text{-}{}^3\mathrm{He})z}=2.7~ \rm fm)|\Psi_{\rm GCM} \rangle=0.75$, $\langle \Phi_{3/2^-_2}(R_{\alpha\text{-}\alpha}=3.5~ {\rm fm}, R_{(2\alpha\text{-}{}^3\mathrm{He})z}=2.6~ \rm fm)|\Psi_{\rm GCM} \rangle=0.80$, and $\langle \Phi_{3/2^-_2}(R_{\alpha\text{-}\alpha}=6.1~ {\rm fm}, R_{(2\alpha\text{-}{}^3\mathrm{He})z}=3.0~ \rm fm)|\Psi_{\rm GCM} \rangle=0.14$, respectively. The maximum overlap appears at $R_{\alpha\text{-}\alpha}=3.5$ fm, indicating that the $3/2_2^-$ state has a slightly expanded spatial configuration compared to the ground state and still maintains a relatively compact shell-model structure.

It is noted that the maximum overlap $\langle \Phi_{3/2^-_3}(R_{\alpha\text{-}\alpha}=6.1~ {\rm fm}, R_{(2\alpha\text{-}{}^3\mathrm{He})z}=2.2~ \rm fm)|\Psi_{\rm GCM} \rangle=0.30$ for the $3/2_3^-$ state is shown in Fig. 4(i).  In Fig. 4 (g) and 4(h), the maximum overlap values are 0.17 and 0.21, respectively. The smaller maximum overlap and the corresponding larger inter-$\alpha$ distance indicate that the obtained $3/2_3^-$ state cannot be approximated by a single cluster configuration and has a dilute gaslike nature.

\begin{figure*}[htbp]
\centering
\includegraphics[width=21cm]{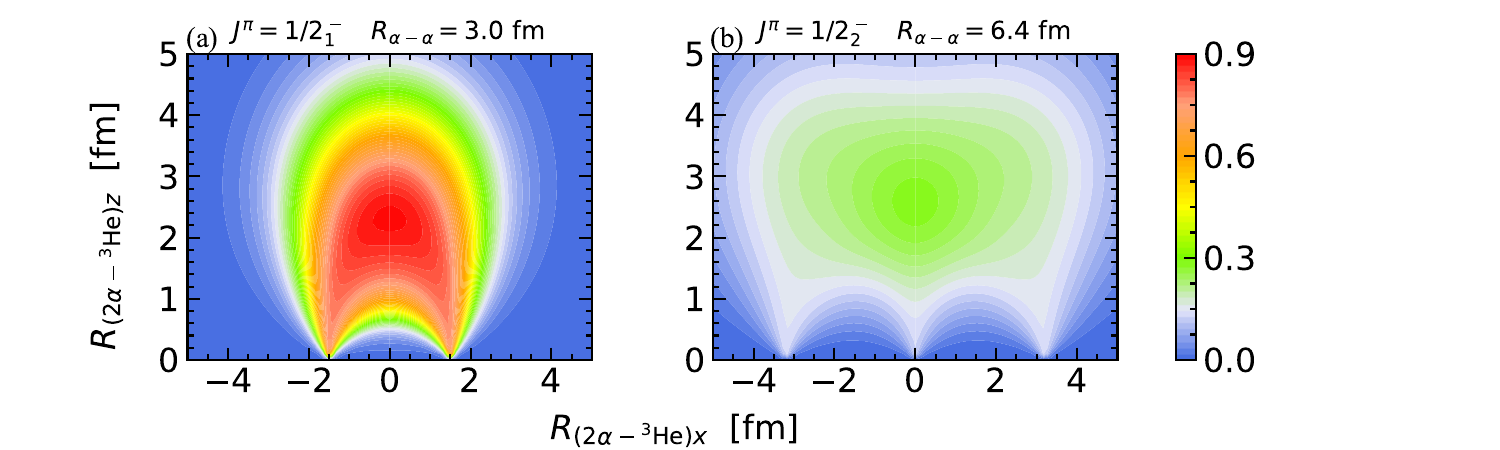}
\caption {Contour plots of the overlap $O(R_{ \alpha\mbox{-}\alpha}, R_{2\alpha\mbox{-}{ }^3 \mathrm{He}})$ for the $J^\pi =1/2^-$ states are displayed as functions of the $x$ component and $z$ component of $R_{ {\rm2\alpha-{}^{3}He}}$. (a) and (b) correspond to the overlaps with the distance $R_{ \alpha-\alpha}$ set at $3.0$ and $6.4$ fm for $1/2_1^-$ and $1/2_2^-$ states, respectively.}
\label{fig5:overlap_neg0.5}
\end{figure*}

\begin{figure*}[htbp]
\centering
\includegraphics[width=21cm]{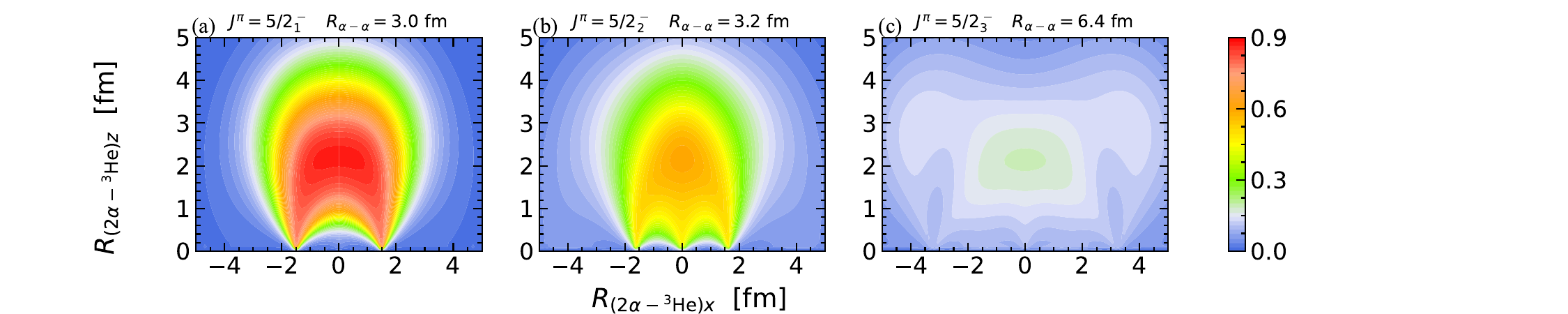}
\caption {Contour plots of the overlap $O(R_{ \alpha\mbox{-}\alpha}, R_{2\alpha\mbox{-}{ }^3 \mathrm{He}})$ for the $J^\pi =5/2^-$ states are displayed as functions of the $x$ component and $z$ component of $R_{ {\rm2\alpha-{}^{3}He}}$. (a)--(c) correspond to the overlaps with the distance $R_{ \alpha-\alpha}$ set at $3.0$, $3.2$, and $6.4$ fm for $5/2_1^-$, $5/2_2^-$, and $5/2_3^-$ states, respectively.}
\label{fig6:overlap_neg2.5}
\end{figure*}

In Fig.~\ref{fig5:overlap_neg0.5} and Fig.~\ref{fig6:overlap_neg2.5}, we present the overlap contour plots of the $1/2^-$ and $5/2^-$ states, respectively. The maximum overlap of $1/2^-_1$ state is $\langle \Phi_{1/2^-_1}(R_{\alpha\text{-}\alpha}=3.0~ {\rm fm}, R_{(2\alpha\text{-}{}^3\mathrm{He})z}=2.4~ \rm fm)|\Psi_{\rm GCM} \rangle=0.89$. For the $1/2^-_2$ state, the maximum overlap is 0.29 at $R_{\alpha\text{-}\alpha}=6.4$ fm and $R_{(2\alpha\text{-}{}^3\mathrm{He})z}=2.6$ fm, indicating a more dispersed spatial configuration due to the larger inter-cluster distance. Similarly, Figs. 6(a)--(c) in Fig.~\ref{fig6:overlap_neg2.5} show the maximum overlap for $5/2_1^-$, $5/2_2^-$ and $5/2_3^-$ states, which are $\langle \Phi_{5/2^-_1}(R_{\alpha\text{-}\alpha}=3.0~ {\rm fm}, R_{(2\alpha\text{-}{}^3\mathrm{He})z}=2.6~ \rm fm)|\Psi_{\rm GCM} \rangle=0.87$, $\langle \Phi_{5/2^-_2}(R_{\alpha\text{-}\alpha}=3.2~ {\rm fm}, R_{(2\alpha\text{-}{}^3\mathrm{He})z}=2.4~ \rm fm)|\Psi_{\rm GCM} \rangle=0.60$, and $\langle \Phi_{5/2^-_3}(R_{\alpha\text{-}\alpha}=6.4~ {\rm fm}, R_{(2\alpha\text{-}{}^3\mathrm{He})z}=2.4~ \rm fm)|\Psi_{\rm GCM} \rangle=0.27$. Moreover, in searching for the maximum overlap, we found another peak for the $5/2_3^-$ state at $R_{\alpha\text{-}\alpha}=2.8$ fm, suggesting the compact structure component. However, the calculated radius and significant enhancement of the ISM transition indicate that the $5/2^-_3$ state may have a clustering nature, as discussed in Ref.~\cite{dell2020experimental}.

\begin{figure*}[htbp]
\centering
\includegraphics[width=21cm]{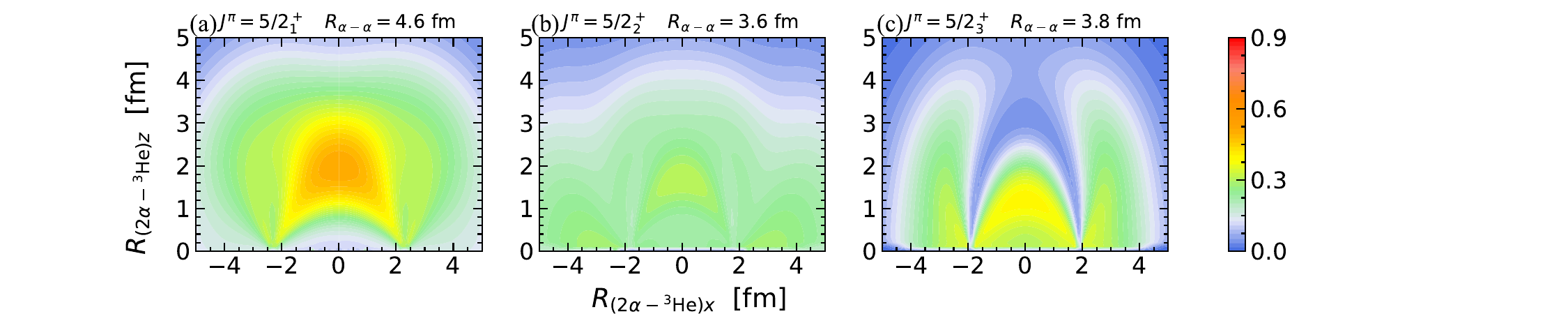}
\caption {Contour plots of the overlap $O(R_{ \alpha\mbox{-}\alpha}, R_{2\alpha\mbox{-}{ }^3 \mathrm{He}})$ for the $J^\pi =5/2^+$ states are displayed as functions of the $x$ component and $z$ component of $R_{ {\rm2\alpha-{}^{3}He}}$. (a)--(c) correspond to the overlaps with the distance $R_{ \alpha-\alpha}$ set at $4.6$, $3.6$, and $3.8$ fm for $5/2_1^+$, $5/2_2^+$, and $5/2_3^+$ states, respectively.}
\label{fig7:overlap_pos2.5}
\end{figure*}

In Fig.~\ref{fig7:overlap_pos2.5}, we present the overlap for the $5/2^+$ states with the corresponding optimal inter-$\alpha$ distances. Figure 7(a) shows the contour plot of the overlap for the $5/2_1^+$ state and the maximum overlap is $\langle \Phi_{5/2^+_1}(R_{\alpha\text{-}\alpha}=4.6~ {\rm fm}, R_{(2\alpha\text{-}{}^3\mathrm{He})z}=2.4~ \rm fm)|\Psi_{\rm GCM} \rangle=0.51$. In Fig. 7(b), the overlap for the $5/2_2^+$ state is $\langle \Phi_{5/2^+_2}(R_{\alpha\text{-}\alpha}=3.6~ {\rm fm}, R_{(2\alpha\text{-}{}^3\mathrm{He})z}=1.8~ \rm fm)|\Psi_{\rm GCM} \rangle=0.31$. 
 For the $5/2_3^+$ state, the maximum overlap is $\langle \Phi_{5/2^+_3}(R_{\alpha\text{-}\alpha}=3.8~ {\rm fm}, R_{(2\alpha\text{-}{}^3\mathrm{He})z}=1.3~ \rm fm)|\Psi_{\rm GCM} \rangle=0.40$. The overlap contour of the $5/2^+_2$ and the $5/2^+_3$ states show extended configurations along the $x$axis. 
In particular, the contour plot for the $5/2^+_3$ state shows larger overlap values at the center and on two sides, suggesting that it may have a linear-chain-like configuration, possibly of the $\alpha-\alpha-\rm {}^{3}He$ or $\alpha-\rm {}^{3}He-\alpha$ type, or a mixed configuration.

From the results of the overlap between the GCM and single projected Brink wave function, we suggest that the $1/2_1^-$, $3/2_1^-$, $5/2_1^-$, $3/2_2^-$, and $5/2_2^-$ states have a compact spatial distribution. The $1/2_2^-$, $3/2_3^-$, and $5/2_3^-$ states have small values of the maximum overlap with large inter-cluster distance, indicating a dilute gaslike nature and suggesting they could be candidates for Hoyle-analog states.

The $^8\text{Be} + {}^{3}\text{He}$ and $^7\text{Be} + \alpha$ thresholds are known to be quite close to the $\alpha + \alpha + {}^{3}\text{He}$ threshold. In fact, the ground states of $^7\text{Be}$ and $^8\text{Be}$ are weakly bound in the $\alpha + {}^{3}\text{He}$ and $\alpha + \alpha$ configurations~\cite{VASILEVSKY200937,ARAI2002963}, respectively. Therefore, examining the $3/2^-$ state by calculating the contributions of the $^8\text{Be} + {}^{3}\text{He}$ and $^7\text{Be} + \alpha$ components may provide insights into the three-body structure.

\begin{figure*}[htbp]
\centering
\includegraphics[width=8cm]{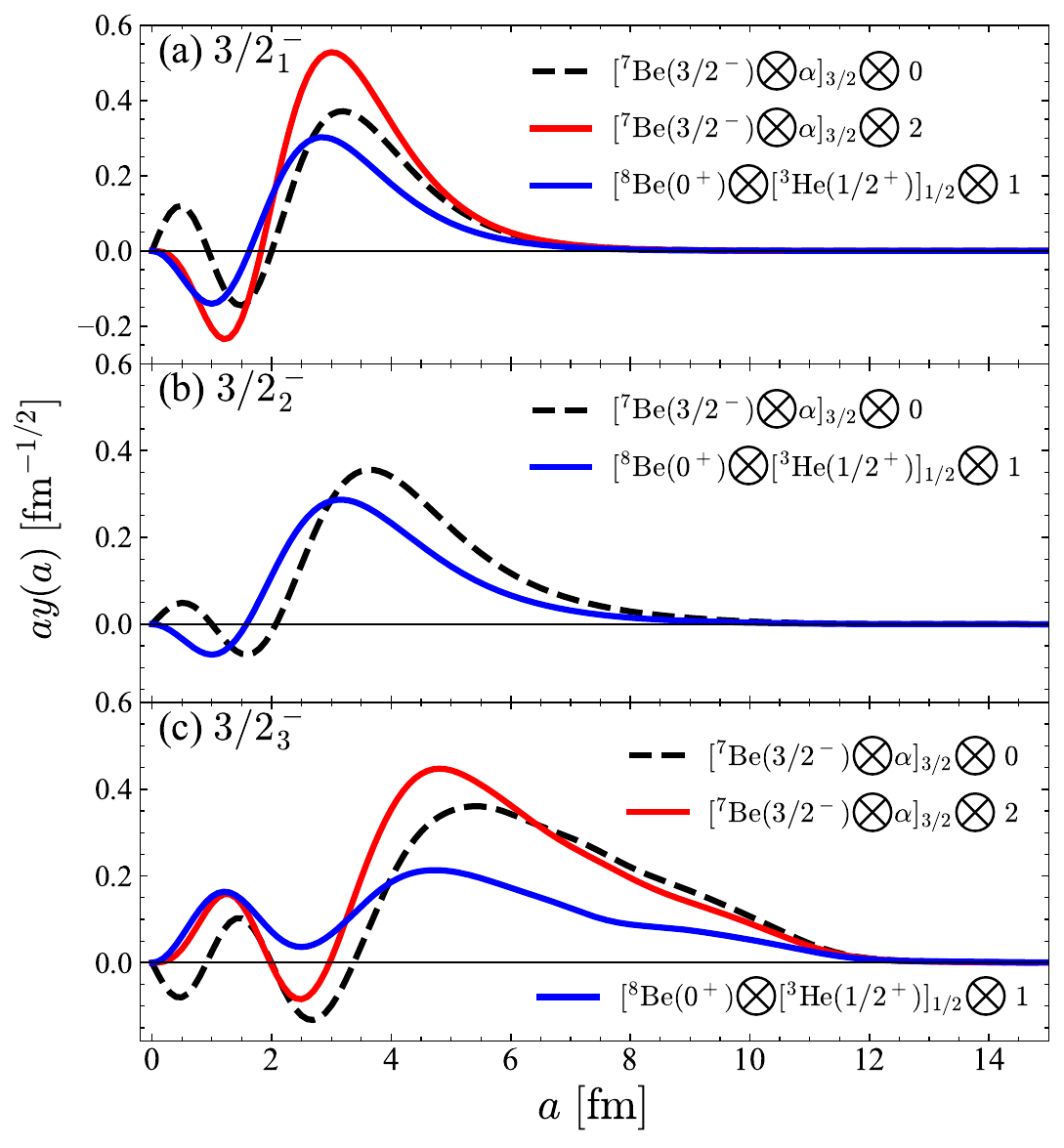}
\caption {The calculated RWAs of the $3/2^-$ states in the $\rm {}^{7}Be+\alpha$ and $\rm {}^{8}Be+{}^{3}He$ channels.  To distinguish the coupling channels, we use the notation $[C_1(j_1) \bigotimes C_2(j_2)]_{j_{12}} \bigotimes l$, where the angular momentum $j_1$ of the $C_1$ cluster and the angular momentum $j_2$ of the $C_2$ cluster couple to form the angular momentum $j_{12}$, and then $j_{12}$ is couple with the orbital angular momentum $l$ of the relative motion between two clusters.  }
\label{fig8:RWAs_neg1.5}
\end{figure*}

To explore the two-body structure of $^{11}\text{C}$, we calculate the reduced width amplitudes (RWAs) for the $^7\text{Be} + \alpha$ and $^8\text{Be}+{}^{3}\text{He}$ channels. As shown in Fig.~\ref{fig8:RWAs_neg1.5}, the channels $^7\text{Be}(3/2^-) + \alpha$ and $^8\text{Be}(0^+) + \rm{}^3He(1/2^+)$ exhibit long tails in the RWAs of the $3/2^-_3$ state, with the latter channel displaying no nodes. This behavior implies a broad spatial distribution, supporting the dilute gaslike nature of the $3/2^-_3$ state. On the other hand, as for other states of $\rm {}^{11}C$, such as $1/2^\pm$, $5/2^\pm$, and $3/2^+$, also have non-negligible contributions from the $^7\text{Be}(3/2^-) + \alpha$ and $^8\text{Be}(0^+) + \rm {}^3He(1/2^+)$ channels. The similarity in these channels likely reflects the underlying $\alpha + {}^{3}\text{He}$ and $2\alpha$ clustering in the low-lying states of $^7\text{Be}$ and $^8\text{Be}$ ~\cite{soic2004alpha, soic2005three}, making it difficult to distinguish between the two-body structures.

\section{summary}
\label{summary}
In this study, we investigate the cluster structure of $\rm {}^{11}C$ by employing three-body GCM calculations, reproducing the energy spectra for both negative-parity and positive-parity states. The r.m.s.\ radius of the $3/2_3^-$ state, together with the enhanced ISM transition strengths from the ground state to the $3/2_3^-$ state, supports the presence of an $\alpha+\alpha+\rm {}^{3}He$ cluster structure in $\rm {}^{11}C$. 
Our analysis of the overlap between GCM and Brink wave functions reveals that the $3/2_3^-$, $5/2_3^-$ and $1/2_2^-$ states show broad distributions, indicative of a dilute gaslike nature. On the other hand, based on the RWA calculations, the $3/2_3^-$ state, in particular, shows long tail in the $\rm {}^{7}Be(3/2^-)+\alpha$ and $\rm {}^{8}Be(0^+)+{}^3He(1/2^+)$ channels, further supporting its dilute gaslike character.  Therefore, we suggest the $3/2_3^-$, $5/2_3^-$, and $1/2_2^-$ state are potential candidates for the Hoyle-analog state, while the $5/2_2^+$ and $5/2_3^+$ states may have a linear-chain-like structure.

\begin{acknowledgments}
B.Z. thanks Professor M. Kimura for valuable discussions that contributed to this work.
This work is supported by the National Key R$\&$D Program of China (2023YFA1606701). This work was supported in part by the National Natural Science Foundation of China under Contract No. 12175042, 11890710, 11890714, 12047514, and 12147101, Guangdong Major Project of Basic and Applied Basic Research No. 2020B0301030008, and China National Key R$\&$D Program No. 2022YFA1602402. This work was partially supported by the 111 Project.
\end{acknowledgments}

%\bibliography{apssamp.bib}% Produces the bibliography via BibTeX.


\end{CJK}
\end{document}